# Vibration-assisted tunneling through single Au adatoms on two-dimensional $WSe_2$

Hitesh Kumar,[1] Yu-Chuan Lin,[2,a] Joshua A. Robinson,[2] and Stefan Fölsch[1,b]

[1]*Paul-Drude-Institut für Festkörperelektronik, Hausvogteiplatz 5-7, Leibniz-Institut im Forschungsverbund Berlin e. V., 10117 Berlin, Germany*

[2]*Dept. Materials Science and Engineering, and Center for 2-Dimensional and Layered Materials, The Pennsylvania State University, University Park, PA 16802, USA*

**ABSTRACT.** Scanning tunneling microscopy (STM) at 5 K was used to study individual Au atoms adsorbed on the surface of a $WSe_2$ layer grown on epitaxial graphene. In line with theoretical predictions, scanning tunneling spectroscopy measurements reveal that the weakly bound adatom gives rise to an electronic state within the energy band gap of the $WSe_2$ layer. Adatoms in different surface locations show different gap-state energy values that follow a random distribution around the Fermi level of the sample with a standard deviation of ~50 meV. The location-dependent shift is attributed to spatial variations in disorder potential. Tunneling via the gap state is accompanied by vibrational excitations as apparent from pronounced sideband peaks in the conductance spectra with Poisson-distributed intensities indicating significant electron-phonon coupling with a Huang-Rhys factor of $S \cong 2.8$. STM tunneling through single Au adatoms on two-dimensional $WSe_2$ constitutes a model case of resonant double-barrier tunneling accompanied by strong coupling to vibrational degrees of freedom.

**Orcid IDs**

H.K.: 0009-0000-4449-0839
Y.-C. L.: 0000-0003-4958-5073
J.A.R.: 0000-0002-1513-7187
S.F.: 0000-0002-3336-2644

---

[a] Present address: Department of Materials Science and Engineering, National Yang Ming Chiao Tung University, Hsinchu City 300093, Taiwan
[b] Corresponding author (foelsch@pdi-berlin.de)

## I. INTRODUCTION

Semiconducting transition metal dichalcogenides (TMDs) [1] are two-dimensional (2D) layered materials occurring in thicknesses down to the single monolayer. Because of their 2D character, TMD layers are susceptible to changes in their properties triggered by atomic point defects like vacancies, dopants, or adsorbates. Consequently, point defects present an important focus of current research dedicated to TMD materials. For example, such defects may lead to single-photon emission [2], tune local electronic properties for semiconductor applications [1,3], or exhibit unique magnetic properties [4,5,6] (see Ref. [7] for a review from the scanning-probe perspective).

Intentional point defects in the form of metal atoms adsorbed on TMD layers were studied before by experiment [5,6,8] and theory [8,9,10,11]. The theoretical studies considered adsorption on monolayer $MoS_2$ and found a significantly stronger bonding of 3d transition metal adatoms (Mn, Fe, Co, Ni) compared to noble metal adatoms (Ag, Au), with the activation energies for surface diffusion following the same trend. It was further predicted that Ag (Au) gives rise to a localized impurity state near the Fermi level mainly deriving from the unpaired Ag 5s (Au 6s) orbital hybridized with the surrounding Mo 4d and S 3p orbitals. These basic properties of individual metal adatoms are important also for the issue of forming electric contacts to TMD layers [12].

Our present work provides experimental evidence for a number of these properties, specifically the emergence of a localized impurity state near the Fermi level. In detail, we performed a cryogenic scanning tunneling microscopy (STM) study of Au atoms adsorbed on a $WSe_2$ layer grown on epitaxial graphene. We show in the following that Au adatoms on $WSe_2$ are weakly bound and prone to random hopping excited by the STM tip. The tip-induced hopping can be exploited to deliberately merge individual adatoms to form $Au_2$ dimers. Most importantly, scanning tunneling spectroscopy (STS) measurements reveal that the discrete Au adatom gives rise to an impurity state situated in the energy band gap of the $WSe_2$ layer. This electronic gap state appears to be coupled to vibrations of the adsorbed atom as apparent from pronounced sidebands in the conductance spectra. The present results exemplify a model case of resonant tunneling through a double barrier accompanied by strong coupling to vibrational degrees of freedom.

## II. EXPERIMENTAL METHODS

The $WSe_2$ layers employed in this work were grown by metal-organic chemical vapor deposition (MOCVD) on epitaxial graphene formed on Si-terminated SiC(0001) following the method described in Ref. [13]. The STM investigations were performed in ultrahigh vacuum (UHV) at a sample base temperature of 5 K. After transfer under ambient conditions, the samples were annealed at 550 K in UHV and loaded into the microscope cooled down to 5 K. Au (99.99% grade purity) was evaporated from an Au droplet held and heated by a tungsten filament. Low coverages (several $10^{12}$ adatoms per $cm^2$) were prepared by dosing directly into the microscope at a sample temperature not exceeding 20 K. STM topography images were recorded in constant-current mode. Bias voltages refer to the sample with respect to the STM tip. STS measurements were performed using a lock-in amplifier at a peak-to-peak modulation voltage of $V_{\text{p-p}}$=5 mV (unless stated otherwise) and a modulation frequency of 675 Hz. We used electrochemically etched tungsten tips cleaned by Ne ion sputtering and electron beam heating.

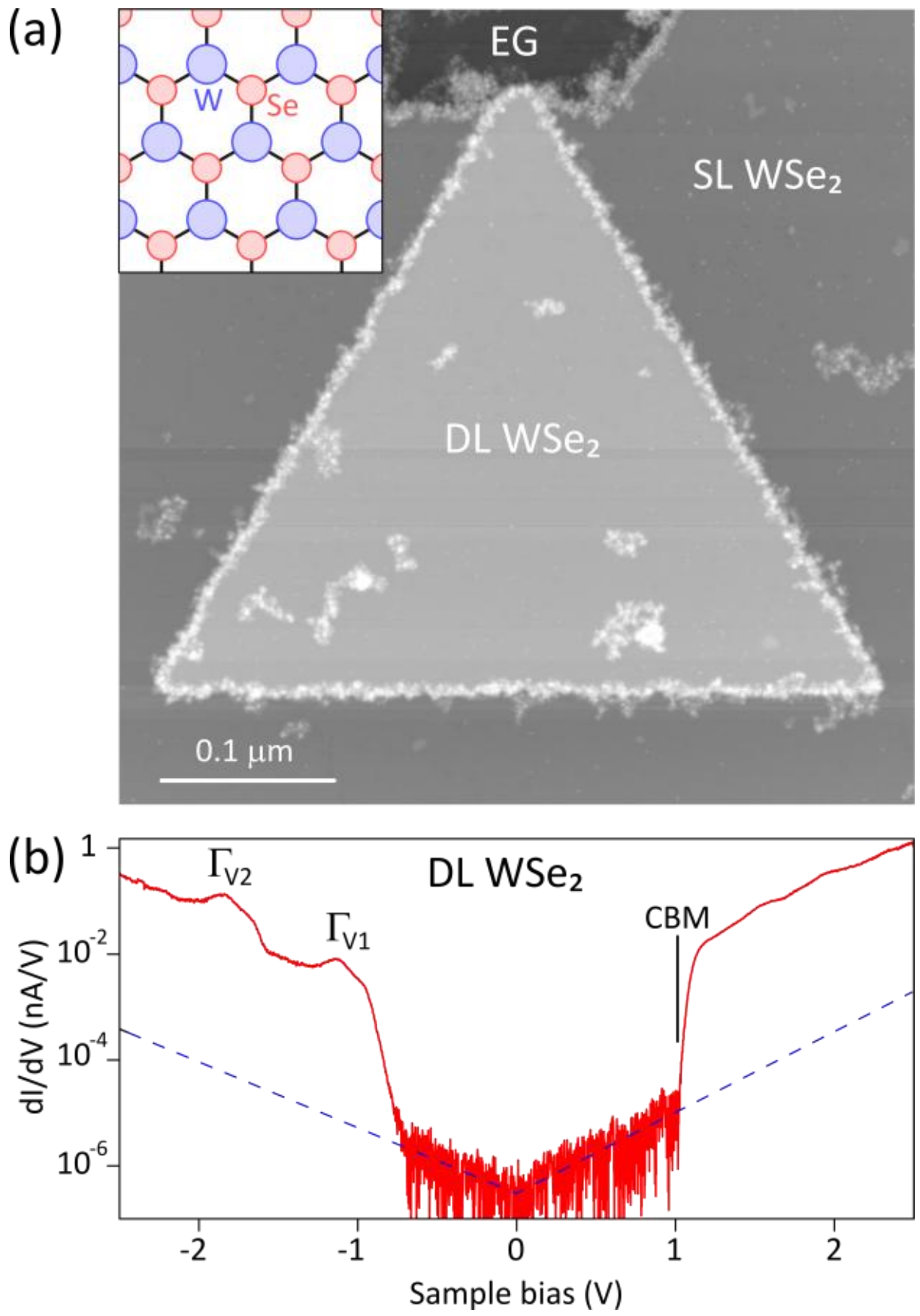

*Fig. 1. (a) Large-scale topography image (5 pA, 3 V) of an extended single-layer $WSe_2$ terrace (SL $WSe_2$) plus a double-layer island (DL $WSe_2$) showing small amounts of residual contamination as well as oxidized step edges; an area of uncovered graphene (EG) is located at the top. (b) Conductance spectrum recorded with the tip probing DL-$WSe_2$ confirming the expected valence-band splitting ($\Gamma_{V1}$, $\Gamma_{V2}$) and a bandgap of ~1.8 V. The spectrum was recorded in variable-z mode (see text for details), the dashed line shows the noise level of the measurement; lock-in modulation $V_{p\text{-}p}$=25 mV.*

## III. RESULTS AND DISCUSSION

In this work, we focus on double-layer $WSe_2$ which is identifiable from its energy bandgap and valence-band splitting due to interlayer coupling [13,14] as revealed by STS measurements of the differential tunnel conductance $dI/dV$ providing a measure of the local surface density of states. Figure 1(a) displays a large-scale topography image of an extended single-layer $WSe_2$ area plus a triangular double-layer $WSe_2$ island with a side length of ~0.4 μm. It is obvious from the image that the step edges are oxidized [15] as a result of exposing the sample to air prior to the STM experiment. The selective oxidization arises because of enhanced reactivity at the Se-terminated zigzag edges, the latter being the step configuration with the lowest formation energy [15]. In contrast, free $WSe_2$ terraces are essentially clean and show only minor residual contamination. The inset shows the corresponding top-view atomic structure of the $WSe_2$ lattice.

Figure 1(b) shows a typical conductance spectrum of double-layer $WSe_2$ recorded over a large bias range covering the valence and conduction band states of $WSe_2$. To record this spectrum, we applied an offset $\Delta s(V)$ to the tip-sample separation $s$ which varied linearly with the magnitude of the sample bias $V$. The exponential increase in conductance arising from this variation in $s$ was then normalized by multiplying the data by a factor of $e^{2\kappa \Delta s(V)}$, where $\kappa$ is an experimentally determined decay constant of 1.1 Å$^{-1}$. This variable-$z$ mode significantly increases the dynamic range of the measurement, facilitating to accurately pin down band-edge positions in STS spectra of semiconductors [16,17]. The spectrum in Fig. 1(b) is consistent with the bandgap (~1.8 eV) and valence-band splitting ($\Gamma_{V1}$ and $\Gamma_{V2}$) expected for double-layer $WSe_2$ [13,14]. Note that all other STS spectra presented below were recorded at *constant* tip-sample separation.

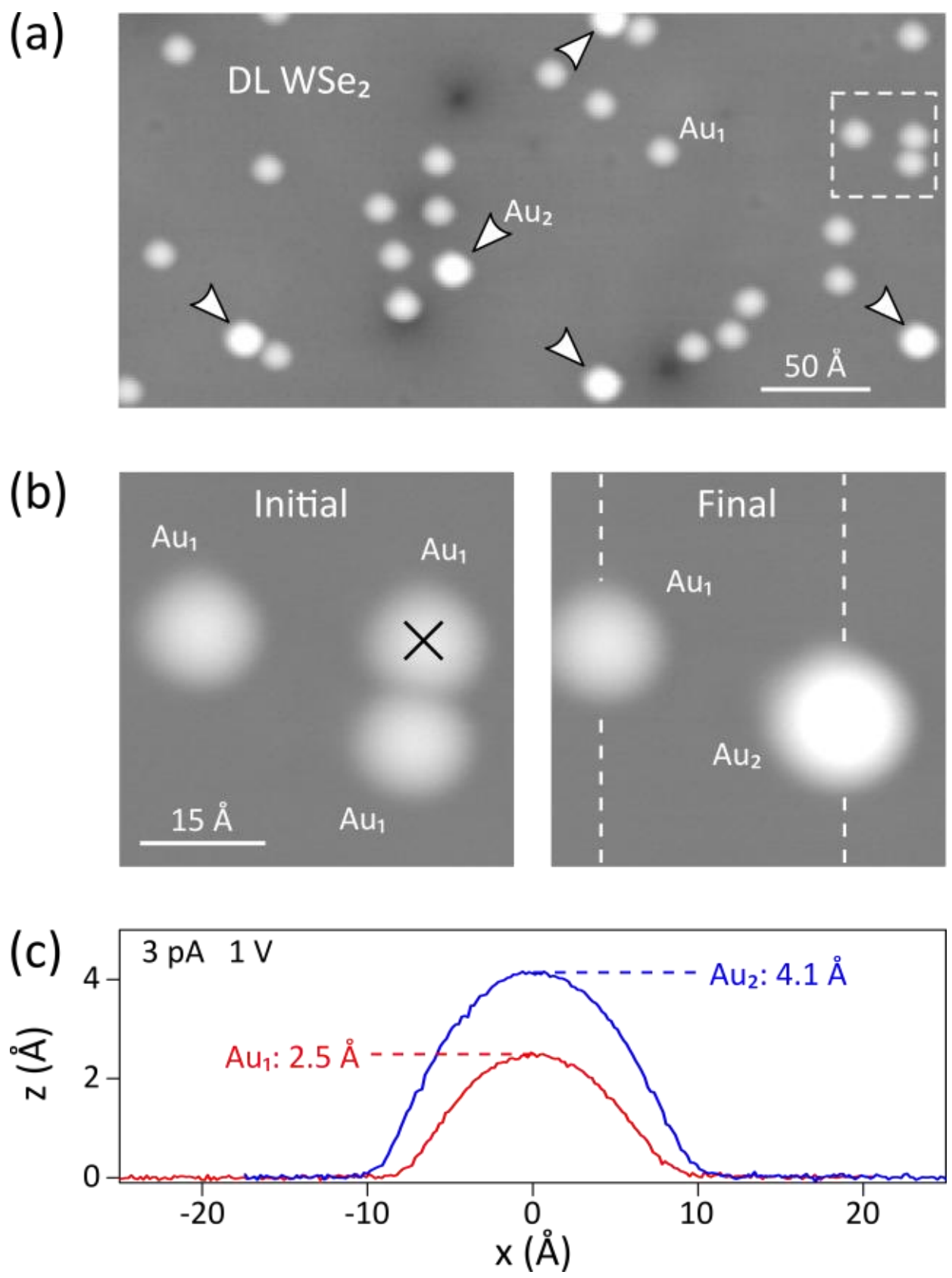

*Fig. 2. (a) STM topography image (368 Å × 180 Å, 3 pA, 1 V) of DL-$WSe_2$ after Au deposition at a sample temperature < 20 K. Lower protrusions are Au adatoms ($Au_1$) and higher protrusions dimers ($Au_2$). Depressions are due to point defects in the DL-$WSe_2$ layer. (b) Left: close-up view of the three adatoms indicated by the dashed box in (a). Right: same area after merging the two nearby adatoms to form a dimer by tunneling at higher current out of the adatom marked in the left image (30 pA, 0.15 V). (c) Constant-current profiles taken at 3 pA and 1 V along the vertical lines in (b) quantifying the apparent height of the adatom (red) and the dimer (blue).*

Au deposition under the aforementioned conditions yields discrete Au adatoms ($Au_1$) and a smaller fraction of dimers ($Au_2$) absorbed on double-layer $WSe_2$ as evident from the constant-current topography image in Fig. 2(a). At the present set-point parameters of 3 pA and 1 V, the dimer is ~1.6 Å taller in apparent height as compared to the adatom. The latter is prone to random hops at higher tunnel currents, indicating a weak adsorptive bonding to the $WSe_2$ surface. This circumstance can be exploited to generate dimers with the STM tip as illustrated by the STM images in close-up view in Fig. 2(b): by positioning the tip above one of the two nearby adatoms at a set-point current of 30 pA (see black cross in the left panel), these two adatoms are merged to form a dimer as seen from the final result shown in the right panel (see the Supplemental Material [18], Fig. S1, for details). Figure 2(c) adds the corresponding constant-current height profiles of the adatom (red) and the dimer (blue), respectively. The apparent height of the tip-generated dimer is the same as for the taller protrusions in panel (a) [marked by arrows] supporting that these are in fact dimers formed directly after deposition. The dimers turn out to be more stable and less susceptible to tip-induced hopping than the adatoms.

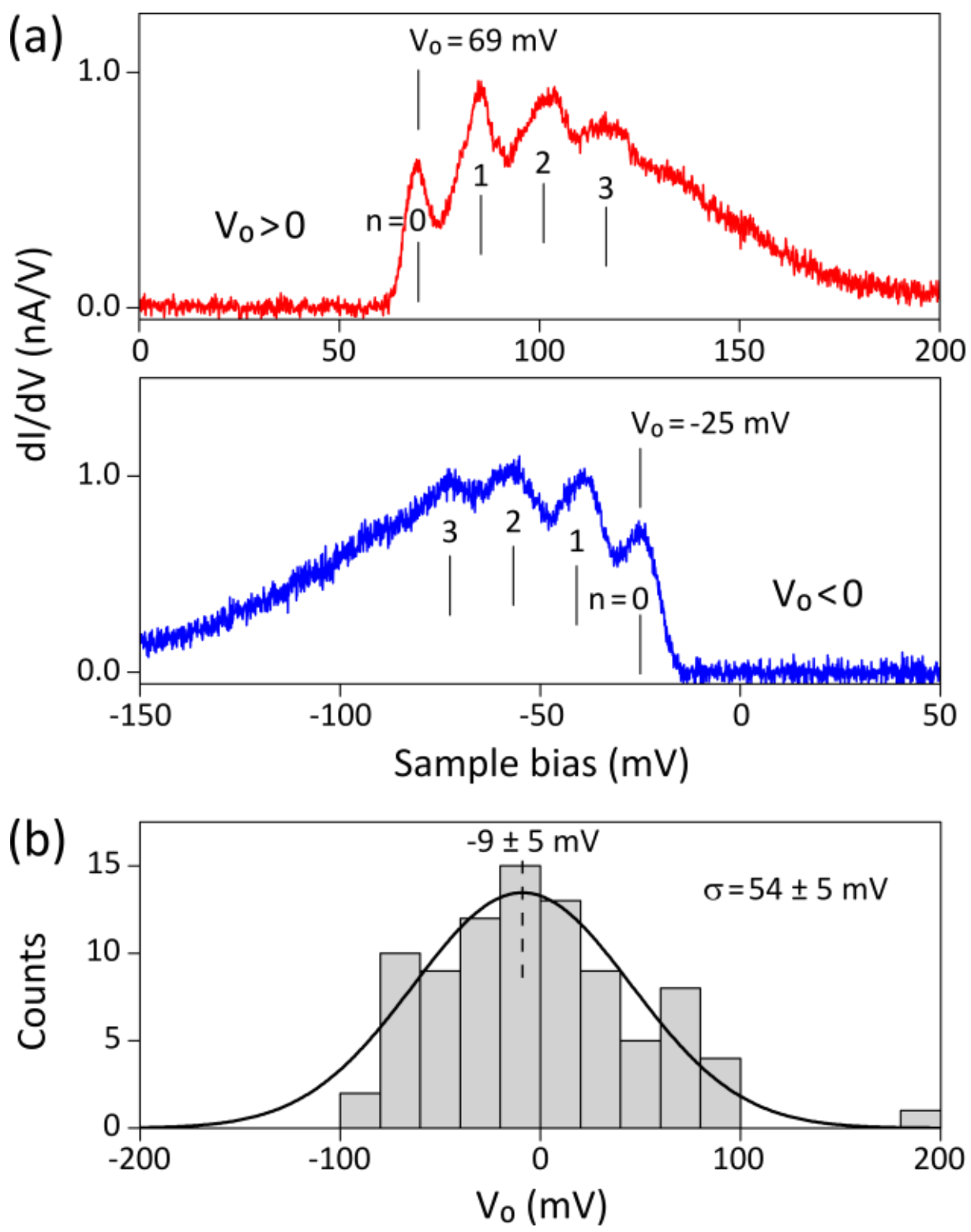

*Fig. 3. (a) Top: conductance spectrum recorded with the tip probing a single Au adatom on DL-$WSe_2$. The peak at $V_0$=69 mV reveals an adatom-induced gap state just above the Fermi level of the sample (at V=0), the sideband peaks arise from vibrational excitations of the adatom upon tunneling via the gap state; n is the excited-state index and n=0 denotes the vibrational ground state. Bottom: spectrum recorded by probing a different Au adatom; the gap state is below the Fermi level and again accompanied by vibrational sidebands. (c) Histogram of conductance-peak positions $V_0$ measured for 88 individual adatoms in different surface locations. The gap-state energy follows a random distribution centered near the Fermi level.*

As noted above, previous theory work predicted that Au atoms adsorbed on monolayer $MoS_2$ give rise to a localized impurity state near the Fermi level and hence in the energy band gap of the TMD layer [8,9,10]. To experimentally verify the existence of an Au-derived gap state in the system studied here, we performed STS measurements at energies within the $WSe_2$ bandgap region. The upper panel of Fig. 3(a) displays a conductance spectrum recorded with the tip held fixed above a discrete Au adatom. It shows a conductance peak at $V_0$= 69 mV revealing the predicted gap state situated just above the Fermi level of the sample (at sample bias $V$=0); this state is accompanied by pronounced sidebands as apparent from the sequence of equally spaced peaks at bias magnitudes > $|V_0|$ with a peak spacing of roughly 16 mV. We attribute the sidebands to the inelastic excitation [19,20] of vibrations of the Au atom in its adsorption potential. Excited states up to $n$=3 are clearly identified as individual peaks where $n$ is the excited-state index and $n$=0 the vibrational ground state. The sideband peaks appear to be successively broadened as $n$ increases; this observation will be further quantified and discussed below.

Strong coupling between localized electronic states and vibrational degrees of freedom was found previously by STS in various different solid-state systems including, for example, single-molecule tunnel junctions [21,22], color centers in ultrathin insulating films [23], single dangling bonds on Si surfaces [24], as well as point defects in TMD layers [4] and in boron nitride [25]. An illustrative approach to describe electron-phonon coupling in charge-carrier transport is the inelastic resonant tunneling model [26] which considers tunneling through an electronic level coupled to a single-frequency harmonic oscillator [27,28,29,30]. The spectrum derived from this model is a progression of equally spaced lines (δ functions) with Poisson-distributed intensities proportional to $e^{-S}(S^n/n!)$ where $S$ is the Huang-Rhys factor quantifying the strength of electron-phonon coupling. To verify this prediction quantitatively, the actual line shape and area of the observed conductance peaks has to be considered. Before turning towards this analysis, we first continue to present the main experimental findings.

The spectrum in the lower panel of Fig. 3(a) was recorded by probing a different adatom in a different surface location. It documents a case where the gap state is just below the Fermi level, in contrast to the situation in the upper panel, as evidenced by the conductance peak at $V_0$=–25 mV. (Notwithstanding this difference, the gap state is again accompanied by vibrational sidebands with an equidistant peak spacing of ~16 mV.) These two contrary cases suggest that the actual energy position of the gap state relative to the Fermi level is governed by electrostatic potential disorder due to residual charged defects – a known effect in semiconductor heterostructures [31,32,33]. Potential disorder is expected to play a significant role also in the system investigated here because of the presence of (typically charged) point defects in MOCVD-grown $WSe_2$ layers (density roughly $10^{12}$ $cm^{-2}$, see Ref. [34] for information on the different point-defect types observed) together with the reduced electrostatic screening in two-dimensional (2D) semiconductors [35]. (Besides this, charge traps have been found to occur also in epitaxial graphene on SiC [36,37].) From a systematic analysis based on a large set of adatoms in various different surface locations we find that the gap-state energy follows a random distribution centered near the Fermi level with a standard deviation of ~50 meV, see Fig. 3(b). In contrast to the case of discrete Au adatoms on $WSe_2$, we have not found any gap state near the Fermi level for $Au_2$ dimers (Supplemental Material [18], Fig. S1).

As shown in Fig. 2(b), Au adatoms on $WSe_2$ are prone to tip-induced hopping. For a given adatom, a hopping event (typically over a distance of several lattice sites) leads to a uniform shift in the corresponding spectrum by a few mV which can be attributed to the spatial variation in disorder potential. In addition, there is indication for temporal changes in disorder potential as well which are likely caused by charge fluctuations associated with nearby point defects as discussed in detail in Appendix A. Taking advantage of the possibility to excite site changes with the STM tip, we analyzed the hopping of Au adatoms with respect to the lattice periodicity of the $WSe_2$ surface. The obtained hopping pattern essentially follows the 1×1 periodicity of the surface lattice showing that the Au adatom favors one specific adsorption site on $WSe_2$ as discussed in Appendix B.

At this point it should be mentioned that a controllable repositioning of Au adatoms by lateral or vertical manipulation with the STM tip [38] turned out to be unfeasible. Attempts to achieve this typically led to random site changes or irreversible atom transfer to the STM tip, suggesting that the binding energy of an Au atom to the tip is significantly larger than to the $WSe_2$ surface.

The data in Fig. 3 indicate that the local disorder potential shifts the Au-induced gap state either into the occupied or the unoccupied regime. Because the gap state is tunnel-coupled to both the STM tip and the sample (the conductive graphene below the $WSe_2$ layer), the present situation can be considered an analogue to the case of a double-barrier tunnel junction [21,23,39,40] with tip and sample acting as "contacts" while the barriers are the $WSe_2$ layer and the vacuum gap between tip and adatom. Upon biasing the junction, voltage drops occur in both barriers: an applied voltage $V_{total}$ is divided into voltage drops $V_{total}\alpha$ and $V_{total}(1-\alpha)$, respectively, with $\alpha<1$. This circumstance enables *bipolar* tunneling because the gap state can be tuned into the energy window between the Fermi levels of the contacts at *both* polarities [21,39,40]. Similar as previously reported for single-molecule tunnel junctions [21] we expect an asymmetric coupling to the contacts ($\alpha\ll1$) because of the large relative permittivity of $WSe_2$ [41] and the fact that the layer thickness is comparable to the size of the vacuum gap, hence resulting in a weaker coupling to the tip than to the sample. Consequently, the tunneling rate for tunneling through the $WSe_2$ barrier is expected to be significantly larger than for tunneling through the vacuum gap.

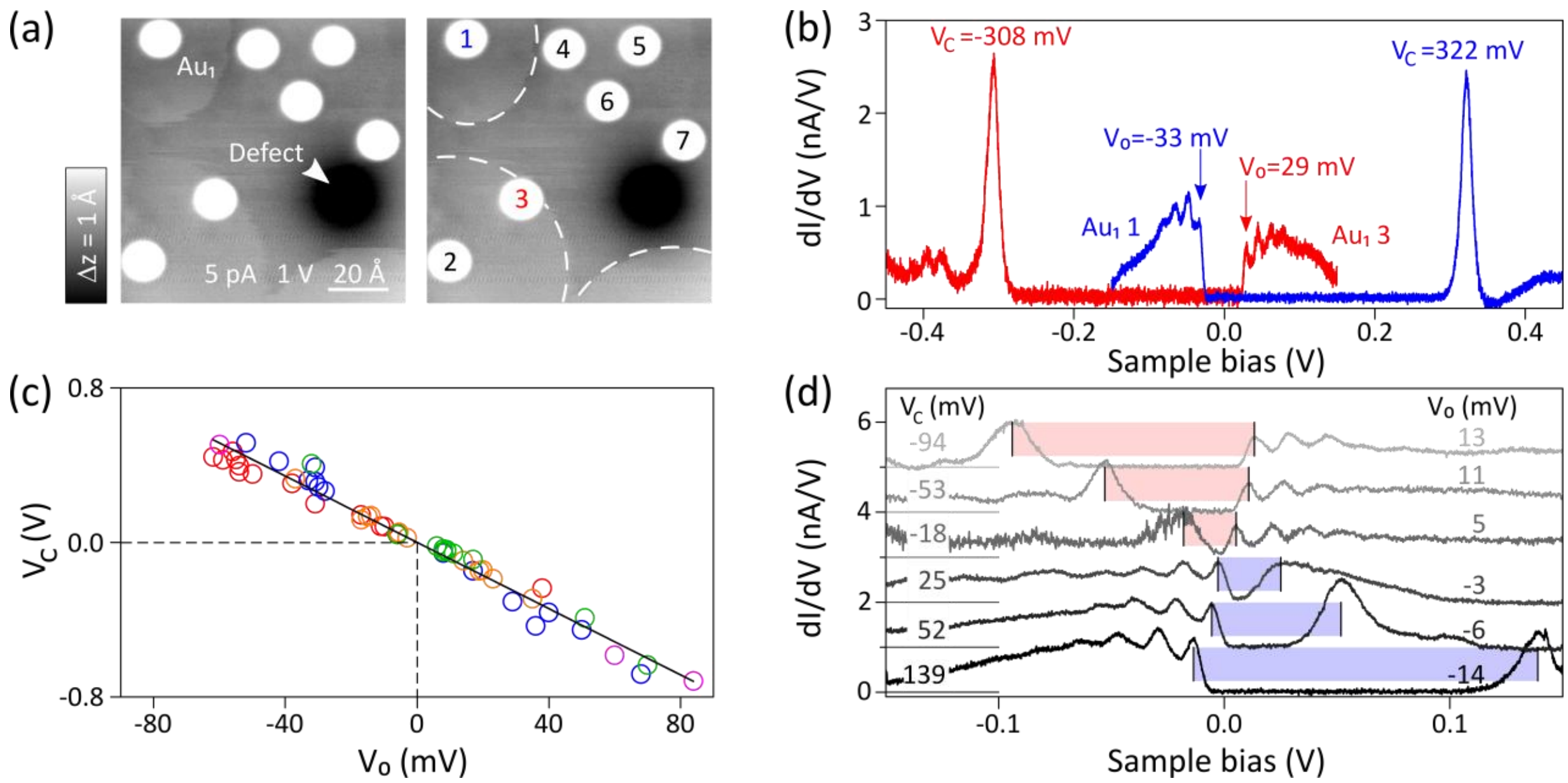

*Fig. 4. (a) Left: topography image (5 pA, 1 V) of seven Au adatoms on DL-$WSe_2$ at enhanced contrast; right: same image with labels and markers discussed in the text. (b) Spectra probing the adatoms labeled 1 (blue) and 3 (red) in (a); aside from the peak at $V_0$ plus sidebands, each spectrum features a pronounced conductance peak at larger bias $V_C$ of opposite polarity arising from bipolar double-barrier tunneling via the gap state. (c) $V_C$ plotted versus $V_0$ confirming a linear correlation of $V_C$ and $V_0$; different colors show measurements in different surface locations. (d) Selection of spectra in which the gap-state energy is tuned from the unoccupied into the occupied states (from top to bottom). The conductance gap $|V_0 - V_C|$ is reduced as the state is brought closer to the Fermi level, see the areas shaded red (blue) in the unoccupied (occupied) states.*

To verify the double-barrier scenario we start with the left hand-side topography image in Fig. 4(a) showing seven adatoms (together with a defect in the $WSe_2$ layer appearing as a depression at the present sample bias of 1 V). The topography is displayed at enhanced contrast to visualize that some of the adatoms are surrounded by disk-shaped features marked by dashed lines in the same image on the right hand-side of panel (a). It is seen that adatoms 1 and 2 *are* surrounded by disks (plus another adatom outside of the image section at the lower right) while adatoms 3 to 7 *are not*.

The origin of the disk-shaped features becomes obvious when recording spectra of larger bias range for adatoms with and without a disk, for example, the adatoms 1 and 3. The blue spectrum shows that adatom 1 gives rise to a conductance peak at negative sample bias $V_0$=−33 mV (plus vibrational sidebands as before) implying that the gap state is occupied when no bias is applied to the junction. On the other hand, at sufficiently large positive bias the gap state is pulled over the Fermi level of the sample so that reversed tunneling can occur from the tip via the gap state and into unoccupied sample states. The blue spectrum reveals the corresponding conductance peak at $V_C$=322 mV (often referred to as the "charging peak" because the adatom is ionized relative to its pristine state). The red spectrum in Fig. 4(b) was recorded on adatom 3 and illustrates the alternative situation: the gap state is now unoccupied ($V_0$=29 mV) and reversed tunneling at opposite bias polarity occurs at $V_C$=−308 mV. The aforementioned disk-shaped features observed in the constant-current STM image in panel (a) are indicative for ionization triggered by the tip-induced electric field, as previously found for near-surface dopants [42] and weakly surface-coupled adsorbates [40,43]. A disk is observed for adatom 1 but not for adatom 3 because only adatom 1 is associated with a charging peak at positive bias. Consequently, upon imaging at even larger positive bias – as in panel (a) – ionization will occur when the scanning tip is sufficiently close to adatom 1. (The event of ionization opens a new tunneling channel and hence increases the apparent height in constant-current imaging [42].)

Consistent with the electrostatics of a double-barrier tunnel junction, Fig. 4(c) confirms the linear correlation of the conductance-peak voltages $V_0$ and $V_C$ as extracted from the spectra of 54 individual adatoms in different surface locations. The mean ratio $|V_C/V_0|$=8.1±1.5 is equivalent to the ratio of voltage drops $(1-\alpha)/\alpha$ in the two barriers so that $\alpha \cong 0.11$, confirming that the tunnel coupling to the contacts is strongly asymmetric in the present case.

Coming back to the spectra in Fig. 4 (b), it is found that the sidebands associated with the conductance peak at $V_0$ are essentially absent at reversed tunneling yielding a single pronounced peak at $V_C$. This observation is in line with previous work on asymmetric double-barrier tunnel junctions [21,30,44] ascribing the suppression of vibrational excitations to rate-limiting processes in the transport through the opaque barrier. The suppression is also obvious from the selection of spectra in Fig. 4(d) in which the gap-state energy was successively tuned from the unoccupied ($V_0$>0) into the occupied regime ($V_0$<0). Starting from the topmost curve (unoccupied), the conductance gap $|V_0-V_C|$ (shaded red) is reduced as the level is brought closer to the Fermi level of the sample. On the other hand, in the occupied regime the conductance gap (shaded blue) is increased as the level is driven away from the Fermi level; this trend is equivalent to the textbook case of a gating potential (here the local disorder potential) acting on a conduction channel (the Au impurity) coupled to external contacts (the tip and the sample, respectively) [39,45].

For a closer analysis of the vibrational sideband spectra, we fitted them by a sequence of lifetime-broadened peaks implying a Lorentzian line shape. In addition, thermal and instrumental broadening (the latter arising from the sinusoidal lock-in modulation) have to be taken into account [46]. To simplify the analysis, both these contributions were approximated by Gaussians yielding a total – thermal plus instrumental – Gaussian broadening with a full width at half maximum of $w_G$=4.6 mV [47]. The spectra were then fitted by a sequence of Lorentzian peaks each convoluted with the Gaussian representing the thermal plus instrumental broadening. (A Lorentzian convoluted with a Gaussian is often referred to as a Voigt profile.) Figure 5(a) illustrates that this procedure yields a good fit of the sidebands as they were observed in the experimental spectra displayed in Fig. 3(a). Convergence of the numerical fit was obtained by allowing the position, the area, and the width of the Lorentzians to vary freely for the levels *n*=0 to 5 [48]. Figure 5(b) reveals that the resulting peak positions are equidistant with a spacing of $\Delta$=16.2±0.2 mV. Taking into account the voltage drop in the $WSe_2$ layer [cf. discussion in context with Fig. 4(c)] this is equivalent to an energy spacing $(1-\alpha)e\Delta$=14.6±0.2 meV of the corresponding vibrational levels. The data collected in panel (b) were obtained from the spectra in panel (a) as well as from three additional spectra recorded on different adatoms (Supplemental Material [18], Fig. S2).

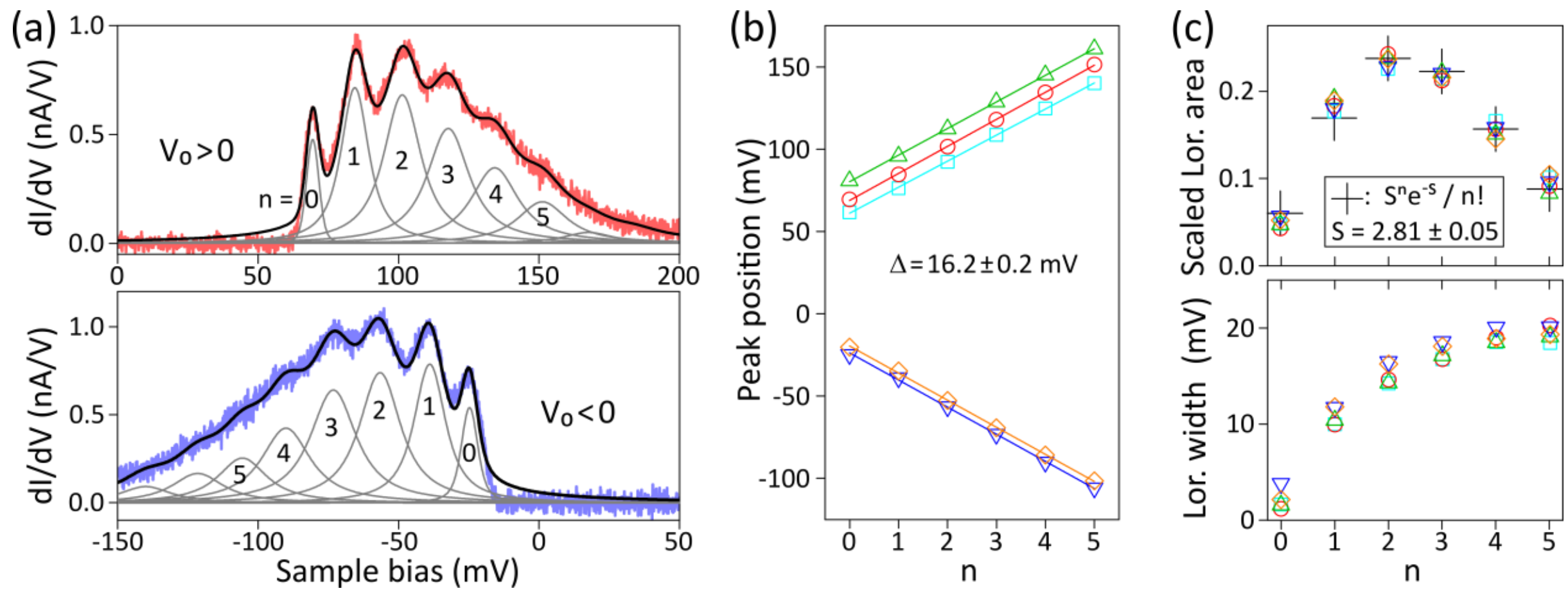

*Fig. 5. (a) Same spectra as in Fig. 3 (red and blue) fitted by a sequence of Lorentzian peaks (gray) each convoluted by a Gaussian to account for thermal and instrumental broadening; the fit (black curve) yields good agreement with the observed sidebands. (b) Peak positions versus n (colored symbols) obtained by fitting five independent spectra showing that the peaks are equally spaced by Δ=16.2±0.2 mV. (c) Top: scaled Lorentzian-peak area versus n (colored symbols) following a Poisson distribution with S=2.81±0.05. Bottom: Lorentzian-peak width (full width at half maximum) showing an increase in linewidth that levels off at larger n.*

The Lorentzian areas derived from each fit are plotted versus *n* in the upper diagram of Fig. 5(c). In this plot, the peak-area values obtained from fitting a given spectrum (colored symbols) were scaled such that they match a Poisson distribution. Indeed, the entire data set is well described by the expression $e^{-S}(S^n/n!)$ with a mean electron-phonon coupling of *S*=2.81±0.05 (black crosses). A Poisson-distributed spectral function implies equilibrated phonons, that is to say, the system fully relaxes to the vibrational ground state between successive electron tunneling events [21,30,44].

The Lorentzian widths (full width at half maximum) as obtained from the spectral fits are plotted versus *n* in the lower diagram of Fig. 5(c). The ground state (*n*=0, no vibration) exhibits a mean width of ~2.4 mV. This broadening is predominantly determined by the tunnel coupling to the sample because the coupling to the tip is significantly weaker, as we have seen in the discussion of Fig. 4. For the excited vibrational states ($n$>0) an increase in linewidth is found which saturates at larger $n$. It can be assumed that the observed linewidth reflects the inverse relaxation time of a given excited state. An increase (saturation) in linewidth would then reflect an increase (saturation) in the number of available relaxation channels. We encourage theoretical modeling of the present double-barrier tunnel junction to pin down these relaxation channels. Existing theory work on molecular transistors [29,30] identified different mechanisms for the broadening of vibrational sidebands: strong tunnel coupling between molecular levels and the density of states of the contacts (finite lifetime of electronic states) as well as dissipation caused by coupling to vibrational modes of the contacts (finite lifetime of phonons). It is likely that a combination of both these mechanisms plays a role in the actual system studied here.

The simple fitting procedure used to analyze the spectra in Fig. 5(a) yields good agreement with the observed sidebands and generates output parameters that are consistent with the picture of a localized electronic state coupled to phonons. But there are also limitations in accurately describing all parts of the conductance spectrum as apparent from the deviation near the onset of the ground-state peak ($n$=0). We expect that a full theoretical modeling of the double-barrier quantum transport would yield an improved description; this modeling, however, is out of the scope of the present experimental paper.

## IV. SUMMARY AND CONCLUSIONS

Cryogenic STM was employed to investigate single Au atoms adsorbed on the 2D semiconductor $WSe_2$ – the latter prepared by MOCVD in the form of double layers on epitaxial graphene formed on Si(0001). The Au adatom is weakly bound to the $WSe_2$ surface so that lateral hops over several lattice sites can be readily excited by the tunneling tip. The resulting hopping pattern indicates that Au favors one specific type of adsorption site. According to previous theory work [9,10,11] the susceptibility to tip-induced hopping as observed here is expected to be less significant for 3d transition metal adatoms; this is consistent with recent STM work reporting stable Fe [5] and Mn adatoms [6] on monolayer $MoS_2$.

Our STS measurements reveal the emergence of an Au-derived impurity state within the $WSe_2$ band gap. Previous theory work [9,10] predicted this gap state to derive from the unpaired Au 6s orbital hybridized with surrounding orbitals of the TMD layer. By probing various different adatoms in different surface locations we found that the gap-state energy follows a random distribution around the Fermi level of the sample with a standard deviation of ~50 meV. We interpret this observation as a manifestation of spatial variations in the local disorder potential felt by the adatom.

The STS data further revealed that the electronic gap state is coupled to phonons giving rise to pronounced vibrational sidebands in the conductance spectra. The sidebands are well fitted by a series of equidistant and lifetime-broadened peaks with Poisson-distributed intensities proportional to $e^{-S}(S^n/n!)$ yielding a mean electron-phonon coupling strength of $S \cong 2.8$. ($S$ values of comparable magnitude were reported for carbon impurities in $WS_2$ layers [4].) The energy spacing of the corresponding vibrational levels deduced from the data amounts to ~15 meV.

Based on the present experimental results, we encourage future theory investigations to gain further insights: for example, a generic modeling of the double-barrier transport could shine light on the observed sideband broadening as a function of $n$. On the other hand, first-principles calculations of the potential energy surface for an Au adatom on $WSe_2$ could clarify the nature of the vibrational mode that is coupled to the Au-derived gap state. More broadly, systematic calculations of the energetics of transition metal adatoms on a TMD layer of choice could guide future experiments with the goal to controllably reposition individual adatoms by means of tip-induced interactions. This could establish a 2D semiconductor platform for exploring the coupling between electronic and vibrational degrees of freedom in engineered nanostructures.

**ACKNOWLEDGEMENTS**

We gratefully acknowledge fruitful discussions with Felix von Oppen. Y.-C.L, and J.A.R. acknowledge funding from NEWLIMITS, a center in nCORE as part of the Semiconductor Research Corporation (SRC) program sponsored by NIST through award number 70NANB17H041 and the Department of Energy (DOE) through award number DE-SC0010697.

## APPENDIX A: SPECTRAL SHIFTS

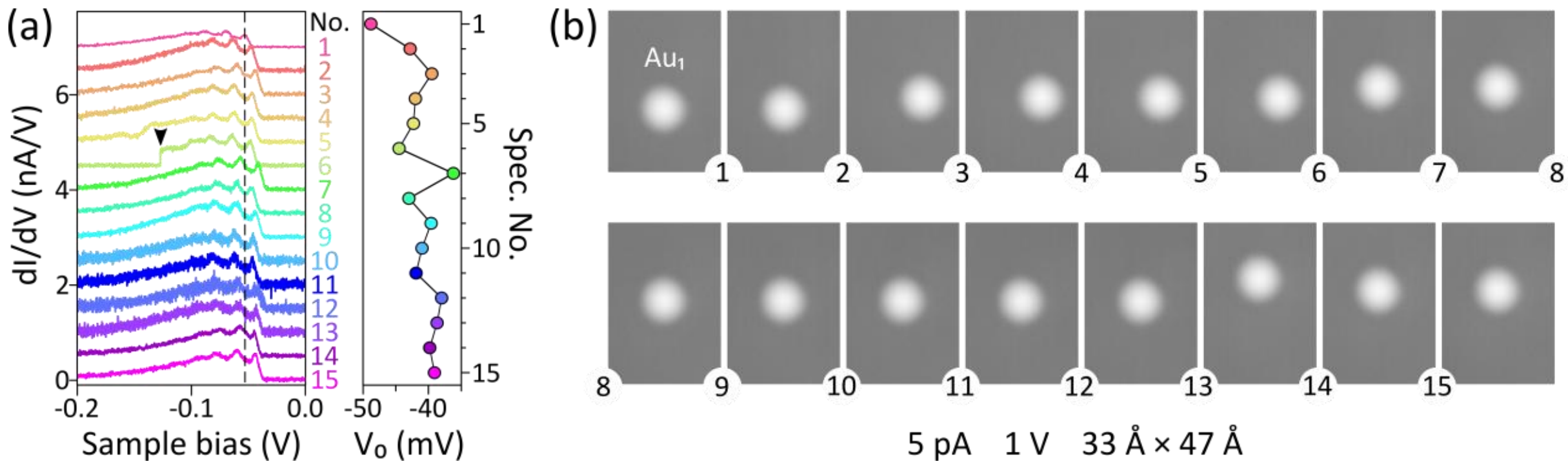

*Fig. 6. (a) Left panel: series of 15 conductance spectra consecutively recorded with the tip probing a discrete Au adatom on DL-WSe$_2$; the spectra are plotted with a vertical offset of 0.5 nA/V for clarity. The gap-state position $V_0$ found in the first spectrum is marked by the dashed vertical line whereas the arrow marks a hopping event while recording spectrum No. 6. Right panel: $V_0$ values as extracted from the spectra. (b) Topography images scanned before and after each spectrum in order to track possible adatom hopping; numbers indicate the order in which the spectra were taken.*

To illustrate the observed spectral shifts discussed in context with Fig. 3, the left-hand side of Fig. 6(a) shows a series of 15 spectra recorded with the tip probing a discrete Au adatom (initial set-point parameters 5 pA and 1 V, bias sweep from 0.50 to −0.25 V for spectra acquisition). It is evident that each spectrum is uniformly shifted relative to the first spectrum (the vertical dashed line marks the gap-state position $V_0$ in the first spectrum as a reference). The actual shift in $V_0$ is quantified in the right-hand side plot of Fig. 6(a) indicating a mean absolute shift of ~3 mV between consecutive spectra.

Before and after each spectrum a constant-current image was taken (same set-point parameters) to track possible site changes occurring during the procedure. The corresponding series of images is displayed in Fig. 6(b) and the order of intermediately recorded spectra is marked by numbers. The images reveal site changes occurring before, while, or after recording spectra 2, 6, 13, and 14: a site change unambiguously occurred while recording spectrum 6 as manifested by the discontinuity at a sample bias of −127 mV marked by an arrow. Regarding the other site changes tracked in the images, it is likely possible that these were triggered while switching the tip set point and position between spectra acquisition and imaging.

Importantly, no site changes occurred in connection with spectra 1, 3 to 5, 7 to 12, and 15 although spectral shifts *do occur* also there as seen in panel (a). This strongly suggests that not only spatial but also temporal changes in local disorder potential are responsible for the shift of the gap-state energy. We speculate that these changes arise from charge traps under the influence of the tip-induced electric field and the carrier transport. Defects acting as charge traps are known to be present in TMDs [49], epitaxial graphene [36,37], and SiC [50].

## APPENDIX B: ADATOM HOPPING

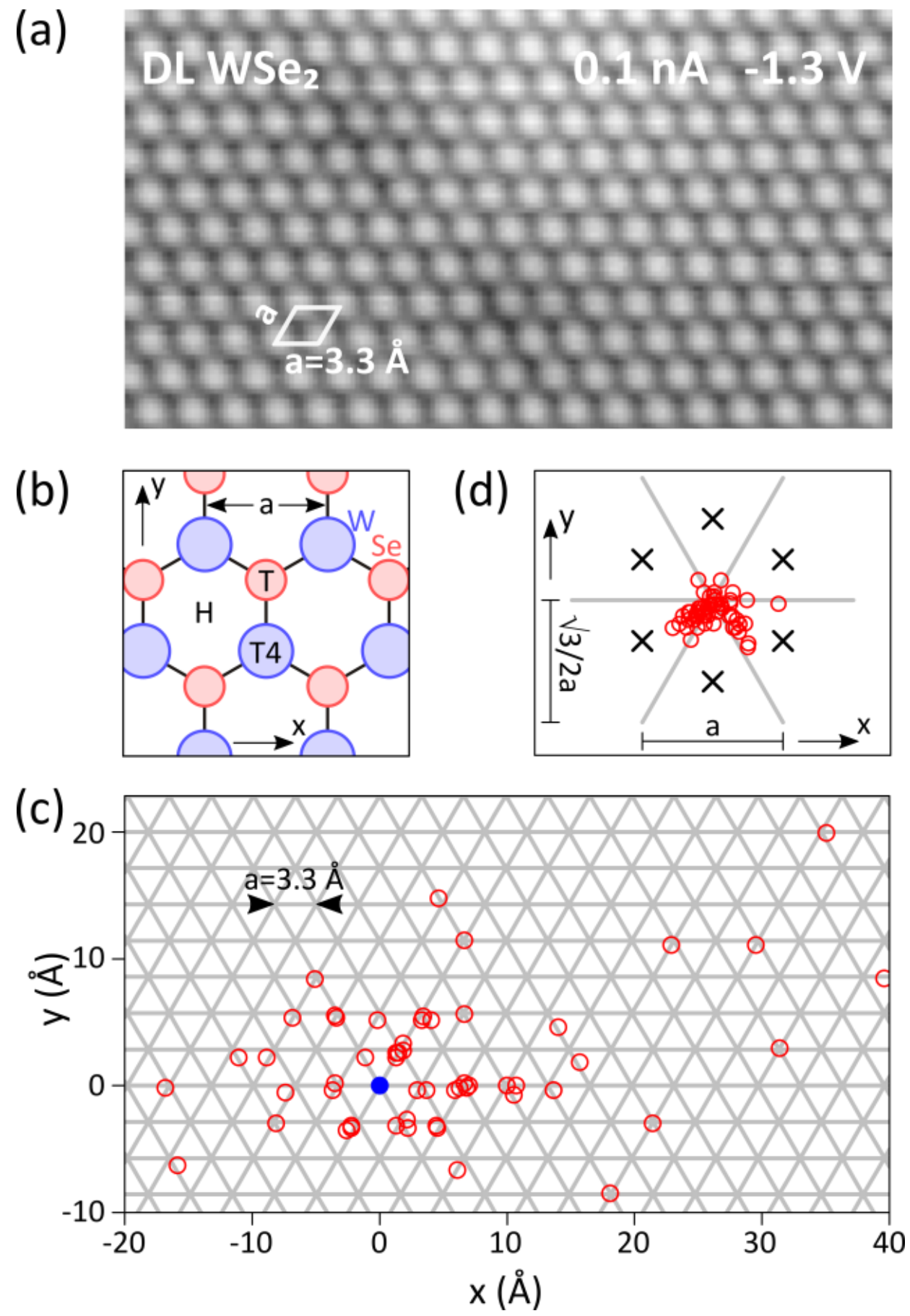

*Fig. 7. (a) Topography image (0.1 nA, -1.3 V, 60 Å × 33 Å) of DL $WSe_2$ at atomic resolution, the unit cell with its lattice constant a=3.3 Å is also shown. (b) Top-view atomic structure of $WSe_2$ showing the different adsorption sites T, T4, and H, each forming a hexagonal sublattice. (c) Sublattice (gray lines) associated with the initial adatom position at x=y=0 (filled blue circle) and final positions (open red circles) after 58 individual hopping events excited by the STM tip. (d) Final positions projected on the sublattice site of the initial position (at the center of the plot), positions belonging to sites of the other sublattices are indicated as small black crosses.*

The hopping pattern of Au adatoms excited by the STM tip reveals preference for one specific adsorption site. To demonstrate this, we start with the topography image in Fig. 7(a) showing the topmost Se sheet of double-layer $WSe_2$ at atomic resolution. This image serves as a reference for the analysis of hopping events discussed below because these cannot be recorded in a controllable way at the set point parameters required to obtained atomic resolution. Theoretical work [8,9,10,11] predicted that metal adatoms can be stable in the on-top site of either the chalcogen (*T*) or the TMD metal (*T4*) or in the hollow site *H*; the bridge site was found to be metastable. Each of the potentially stable adsorption sites forms a hexagonal sublattice with a lattice constant of *a*=3.3 Å, see the scheme in Fig. 7(b).

The hexagonal grid in Fig. 7(c) represents such a sublattice; the area covered by this grid has the same size as the topography image in panel (a). We use this grid to visualize the hopping pattern obtained from 58 individual hopping events of adatoms on double-layer $WSe_2$ islands in four macroscopically different surface locations. The initial adatom position of each hopping event is marked by the filled blue circle at *x*=*y*=0 while the open red circles indicate the final positions. It is evident that the final positions fall on the grid representing the sublattice of the initial adsorption site. This is seen in more detail in panel (d) showing the final positions projected on the sublattice site of the initial position (at the center of the plot). Small black crosses mark positions belonging to sites of the other sublattices. It is clear from this plot that transitions between sublattices are unfavorable upon tip-induced hopping. Based on existing theory predictions for Au adatoms bound to $MoS_2$ [9,10,11], it is likely that Au shows a similar preference for adsorption in the *T* site on top of surface Se in the present case.

[47] The thermal broadening is $3.53kT$ while the instrumental broadening amounts to $1.73V_p$ with $V_p$ the amplitude of the sinusoidal modulation. The total broadening is then approximated by a Gaussian having a full width at half maximum of $w_G = \sqrt{(3.53kT/e)^2 + (1.73V_p)^2}$ yielding $w_G$=4.6 mV under the experimental conditions $T$=5 K and $V_p$=2.5 mV.

[48] The peaks corresponding to levels $n$>5 were pinned to equidistant positions and their width was set to the value obtained from the fit for $n$=5; the only free parameter left for $n$>5 was the peak area.

# Supplemental Material to:

## Vibration-assisted tunneling through single Au adatoms on two-dimensional $WSe_2$

Hitesh Kumar,[1] Yu-Chuan Lin,[2,a] Joshua A. Robinson,[2] and Stefan Fölsch[1,b]

[1]*Paul-Drude-Institut für Festkörperelektronik, Hausvogteiplatz 5-7, Leibniz-Institut im Forschungsverbund Berlin e. V., 10117 Berlin, Germany*

[2]*Dept. Materials Science and Engineering, and Center for 2-Dimensional and Layered Materials, The Pennsylvania State University, University Park, PA 16802*

**Contents:**

[a] Present address: Department of Materials Science and Engineering, National Yang Ming Chiao Tung University, Hsinchu City 300093, Taiwan

[b] Corresponding author (foelsch@pdi-berlin.de)

## S1. Tip-induced $Au_2$ dimer formation

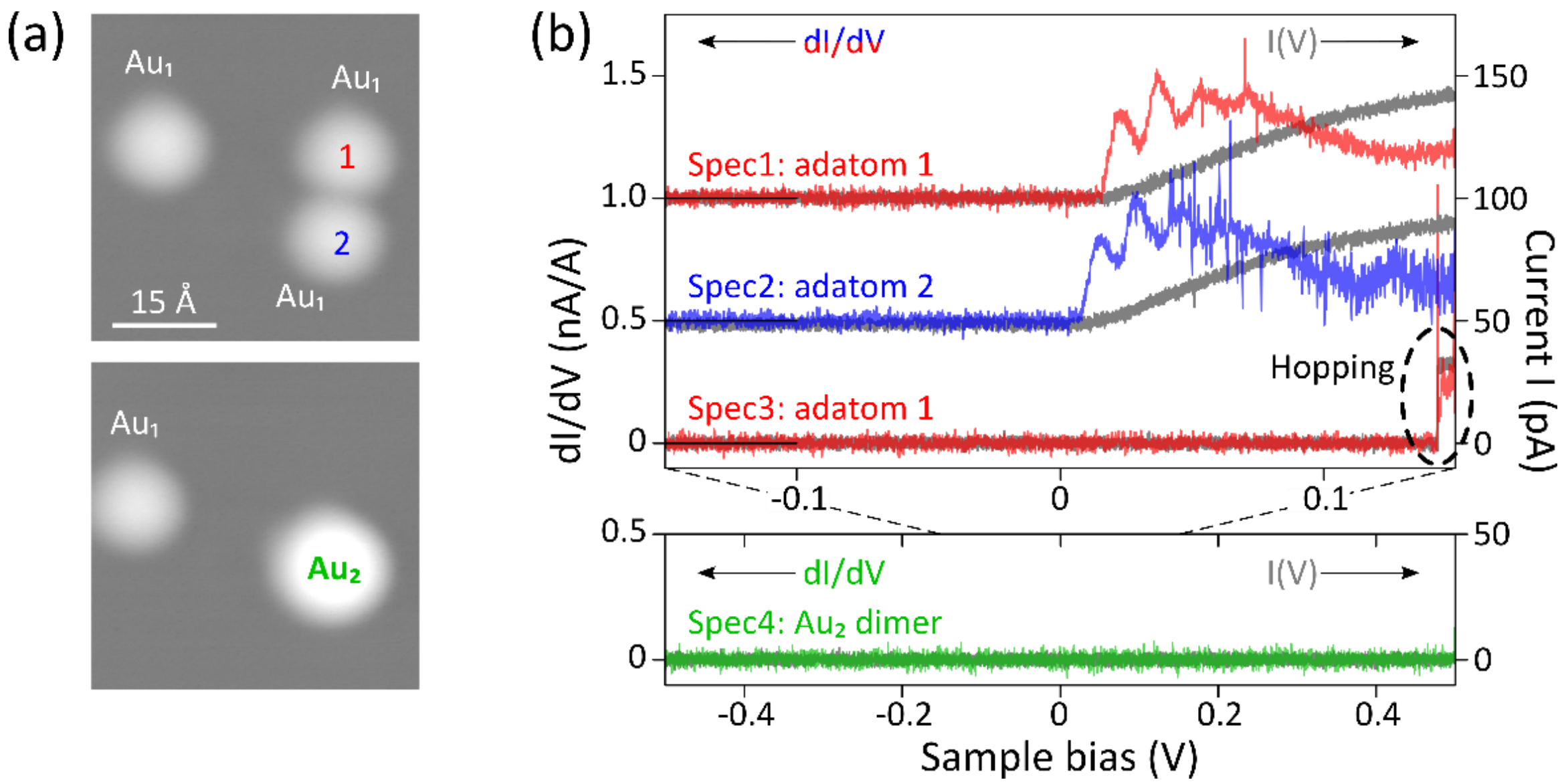

**Fig. S1.** (a) Topography images (3 pA, 1 V) as in Fig. 2(b) of the main text showing the adatoms labeled 1 and 2 prior to tip-induced dimer formation (top) and the resulting dimer (bottom). (b) Conductance spectra (red, blue, and green) recorded prior to (Spec1 & Spec2), during (Spec3), and after the dimer formation (Spec4); corresponding *I(V)* curves are shown in gray. For clarity, spectra in the upper panel are offset along the y axes (left axis: *dI/dV*, right axis: current *I*). The discontinuity in Spec3 (marked by the thick dashed line) indicates the tip-induced adatom hopping and dimer formation.

The conductance spectra in Fig. S1(b) were recorded prior to (Spec1 and Spec2), during (Spec3), and after the dimer formation (Spec4) discussed in context with Fig. 2(b) in the main text; the same images shown there are reproduced in panel (a). Colored curves show the signal of the *dI/dV* channel for each spectrum, the corresponding *I(V)* channel is plotted in gray. The first two spectra Spec1 and Spec2 highlight the impurity state (plus vibrational sidebands) associated with Au adatom 1 and 2 while the second spectrum taken on adatom 1 (Spec3) indicates the hopping and dimer formation at the beginning of the bias sweep starting at 0.15 V, see the discontinuity in signal marked by the thick dashed line. The spectrum in the lower panel was recorded on the resulting $Au_2$ dimer (Spec4) and reveals that no dimer-derived state is detected within the probed gap region.

## S2. Auxiliary sideband spectra

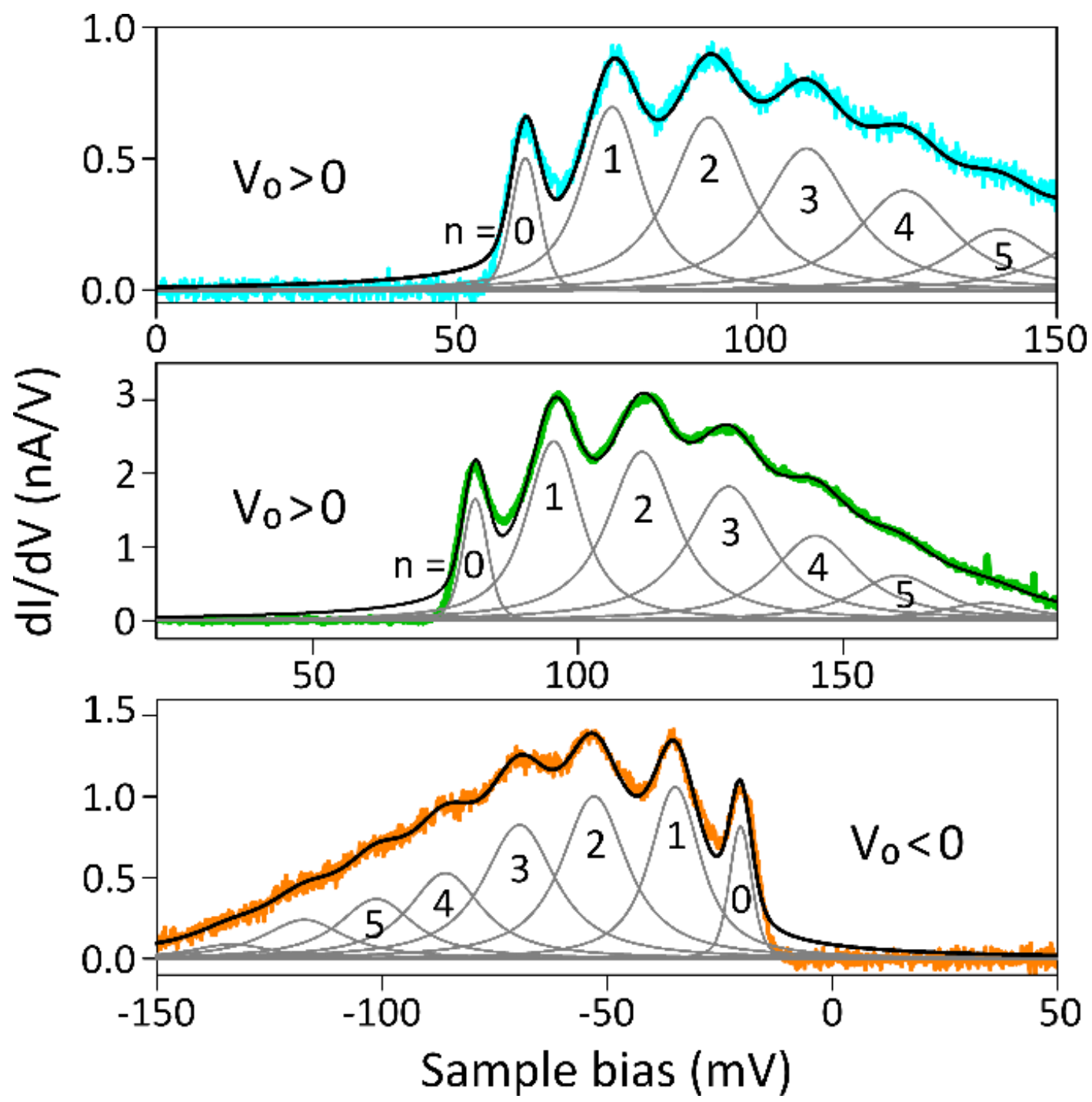

**Fig. S2.** Auxiliary conductance spectra (colored curves) each taken on a single Au adatom in a different surface location. The sideband fits (black curves) were carried out as discussed in context with Fig. 5 in the main text. The positions, scaled areas and widths of the Voigt profiles (gray curves) as a function of the excited-state index *n* are included as colored symbols in Figs. 5(b) and (c) in the main text.